\documentclass[a4paper,conference]{IEEEtran}
\usepackage[a4paper, left=0.514in, right=0.514in, bottom=1.569in, top=0.75in]{geometry}
\usepackage{adjustbox}
\usepackage{cite}
\usepackage{amsmath,amssymb,amsfonts}
\usepackage{algorithm,algorithmic}
\usepackage{graphicx}
\usepackage{textcomp}
\usepackage{hyperref}
\usepackage{soul}
\usepackage{color}
\usepackage{multirow}
\usepackage{booktabs}
\usepackage{threeparttable}

\ifCLASSINFOpdf
\else
\fi
\begin{document}
%
\title{Attention2AngioGAN: Synthesizing Fluorescein Angiography from Retinal Fundus Images using Generative Adversarial Networks}

\author{\IEEEauthorblockN{Sharif Amit Kamran\IEEEauthorrefmark{1}, Khondker Fariha Hossain\IEEEauthorrefmark{2}, Alireza Tavakkoli\IEEEauthorrefmark{3} and Stewart Lee Zuckerbrod\IEEEauthorrefmark{4}}
\IEEEauthorblockA{\IEEEauthorrefmark{1}\IEEEauthorrefmark{3}
\textit{University of Nevada, Reno}
NV, USA\\
\IEEEauthorrefmark{2}
\textit{Deakin University}
Melbourne, Australia\\
\IEEEauthorrefmark{4}
\textit{Houston Eye Associates}
Houston, TX, USA\\
skamran@nevada.unr.edu\IEEEauthorrefmark{1}, khossain@deakin.edu.au\IEEEauthorrefmark{2}, tavakkol@unr.edu\IEEEauthorrefmark{3}, szuckerbrod@houstoneye.com\IEEEauthorrefmark{4}}
}


%


\maketitle

\begin{abstract}
Fluorescein Angiography (FA) is a technique that employs the designated camera for Fundus photography incorporating excitation and barrier filters. FA also requires fluorescein dye that is injected intravenously, which might cause adverse effects ranging from nausea, vomiting to even fatal anaphylaxis. Currently, no other fast and non-invasive technique exists that can generate FA without coupling with Fundus photography. To eradicate the need for an invasive FA extraction procedure, we introduce an Attention-based Generative network that can synthesize Fluorescein Angiography from Fundus images. The proposed gan incorporates multiple attention based skip connections in generators and comprises novel residual blocks for both generators and discriminators. It utilizes reconstruction, feature-matching, and perceptual loss along with adversarial training to produces realistic Angiograms that is hard for experts to distinguish from real ones. Our experiments confirm that the proposed architecture surpasses recent state-of-the-art generative networks for fundus-to-angio translation task. 
\end{abstract}



%
\IEEEpeerreviewmaketitle

\section{Introduction}

Retinal Funduscopy along with Fluorescein Angiography (FA) has been a popular diagnosing tool for retinal vascular and adverse pigment epithelial-choroidal conditions \cite{mary2016retinal}. In Fluorescein Angiography, a fluorescent dye is injected in the optic vein. It becomes noticeable after 10 minutes of insertion, depending on the age and the cardiovascular structure of the retinal layers \cite{mandava2004fluorescein}. While commonly considered healthy, non-fatal complications can arise, such as allergic reactions, nausea, vomiting, etc. Moreover, fatal cases have been documented, with symptoms such as anaphylaxis, heart attack, anaphylactic shock due to the leakage of the dye in the intravenous space \cite{lira2007adverse,kwan2006fluorescein,kwiterovich1991frequency}.

Many automated systems have been proposed for the diagnosis of intrinsic conditions and diseases from fundus photos. They generally comprise of different image processing techniques and machine learning algorithms \cite{gurudath2014machine,fu2018disc,poplin2018prediction,lira2007adverse}.  Currently, there is no computational inexpensive alternative for generating reliable and reproducible fluorescein angiography images. Retinal fundoscopy is the only alternative for differential diagnosis that is easily available and financially viable. Optical coherence Tomography combined with basic image processing\cite{opticnet19} can be utilized for the diagnosis of retinal disease but is too expensive and not widely accessible in developing economies. As it stands, an efficient and faster procedure is crucial for avoiding any potential hazards associated with invasive fluorescein angiography.

In this paper, we introduce Attention-AngioGAN, a robust conditional Generative Adversarial Network (GAN) to produce fluorescein angiograms from retinal fundus images. For qualitative evaluation, we compare our automated technique with recent state-of-the-art GAN architectures such as pix2pixHD~\cite{wang2018high}, U-GAT-IT~\cite{kim2019u}, and Stargan-V2~\cite{choi2020stargan}. Moreover, we used Frechet inception Distance (FID)~\cite{heusel2017gans} and Kernel Inception Distance (KID)~\cite{binkowski2018demystifying} scores to quantify image quality and calculate similarity with real angiograms.

\begin{figure*}[tp]
    \centering
    \includegraphics[width=16cm,height=9cm]{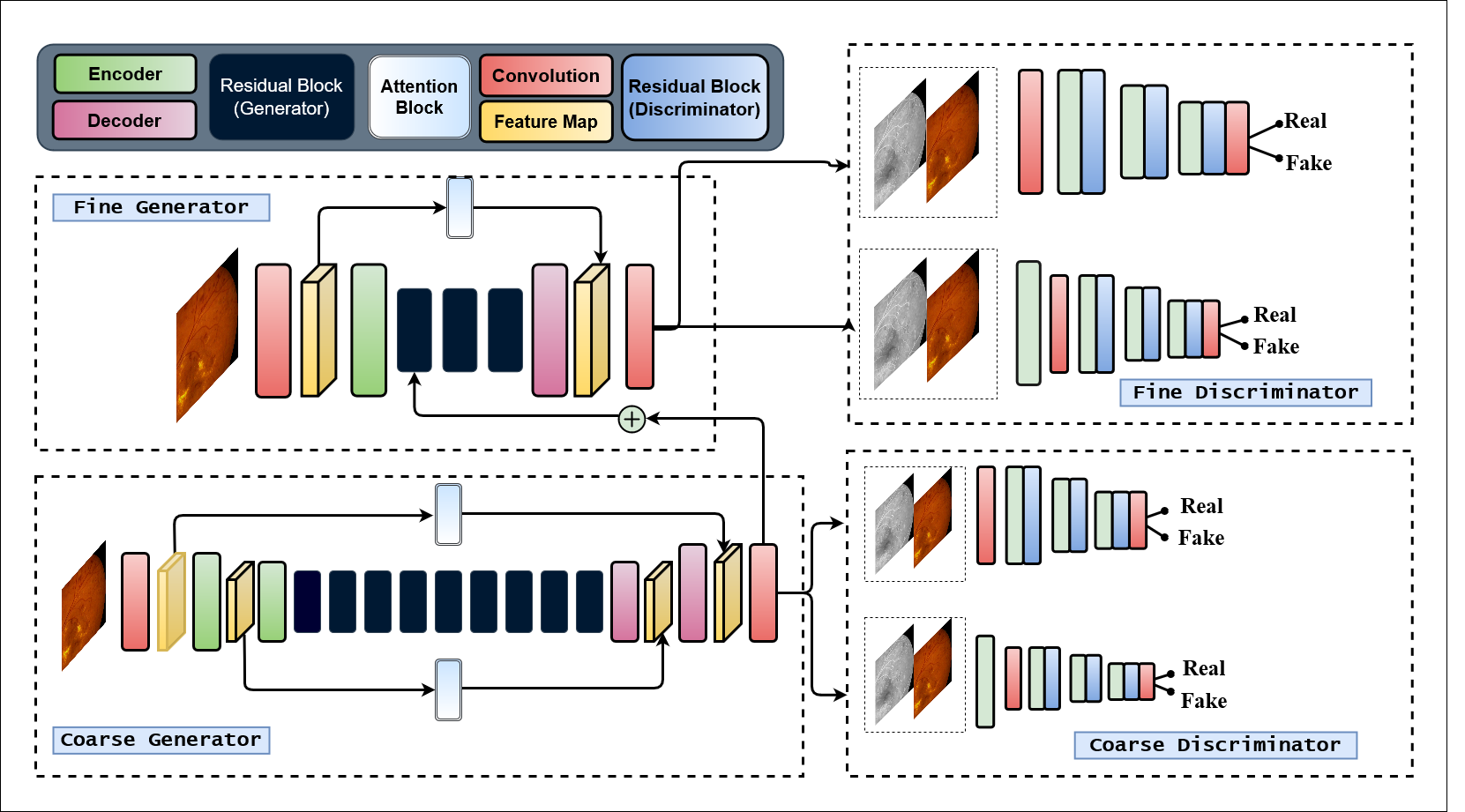}
    \caption{The Proposed GAN architecture consist of two Generators [one Fine and one Coarse], four Discriminators [two Fine and two Coarse]. The fine discriminators take input image as the sample size while the coarse discriminators take input as half of the sample size.}
    \label{fig1}
\end{figure*}

\section{Literature Review}



Recently, there has been a surge of Generative adversarial networks(GAN) based applications ranging from image translation \cite{chen2018sketchygan,sangkloy2017scribbler}, editing \cite{zhu2016generative,dekel2018sparse} and image style transfer \cite{wang2018high,xian2018texturegan}. GANs can potentially extract and learn fine and coarse information from images by combining multiple architectures having multi-scale resolution \cite{burt1983laplacian,brown2003recognising}.
Such examples are wide-spread in both Conditional \cite{huang2017stacked,denton2015deep} and Unconditional  GAN settings\cite{chen2017photographic,zhang2017stackgan}. By incorporating multiple high-resolution architectures, they can learn distinct domain-specific features with high precision and robustness.

Emphasizing image to image translation, numerous prior work has been proposed, where they focused on architectural changes to acquire a higher quality result. To illustrate, pix2pixHD~\cite{wang2018high} utilized  PatchGAN as a multi-scale discriminator to achieve better visual representation containing local and global information. On the other hand,  U-GAT-IT~\cite{kim2019u}, an unsupervised learning architecture, extracts local features and texture by incorporating AdaLIN (Adaptive Layer-Instance Normalization). While most architectures ensure high quality of images, Stargan-V2~\cite{choi2020stargan}, a style-transfer network, focuses on the domain-specific features. By doing so, they allow diverse and scalable image-to-image translation in multiple domains within a single model.

Most of the image-to-image translation models are either focused on domain level transformation or combining style and textures of two images. For instance, high-resolution images generated by U-GAT-IT~\cite{kim2019u} and Stargan-V2~\cite{choi2020stargan} use attention modules to extract local features information and don't utilize perceptual loss. Whereas StyleGAN \cite{park2019semantic}, pix2pixHD  \cite{wang2018high} emphasizes more on incorporate perceptual loss with different styles of target images. By incorporating both these ideas we propose an architecture, where we combine perceptual loss and multi-scale discriminator to retain global information like the shape of optic-disc, color, contrast, etc. On the other hand, we utilize feature matching loss and introduce new multi-attention modules to retain local features like retinal venules, arteries, protein buildup, and microaneurysm. The visual representation and quantitative result prove that our proposed technique surpasses state-of-the-architectures and tricks, expert ophthalmologists, to think they are authentic.


\section{The Proposed Methodology}
This paper proposes an attention-based generative adversarial network (GAN) comprising of separate new residual blocks for generators and discriminators. Moreover, the training includes perceptual, feature matching, and reconstruction loss to synthesize more vivid looking angiograms from retinal fundus images. First, we discuss about coarse and fine generators in section \ref{subsec:generators}. Next, we elaborate our building blocks in Section \ref{subsec:encdec}, \ref{subsec:residualblock}, \ref{subsec:attention}. We then delve into the multi-scale discriminators and their interconnection with the generators to define the whole end-to-end pipeline for the generative network in sections  \ref{subsec:discriminators}. Ultimately, in section \ref{subsec:objective}, we discuss the loss minimizing and maximizing function and loss weight distributions for different losses interrelated to each of the separate architecture that forms the proposed model.

\begin{figure}[htb]
    \centering
    \includegraphics[width=8cm]{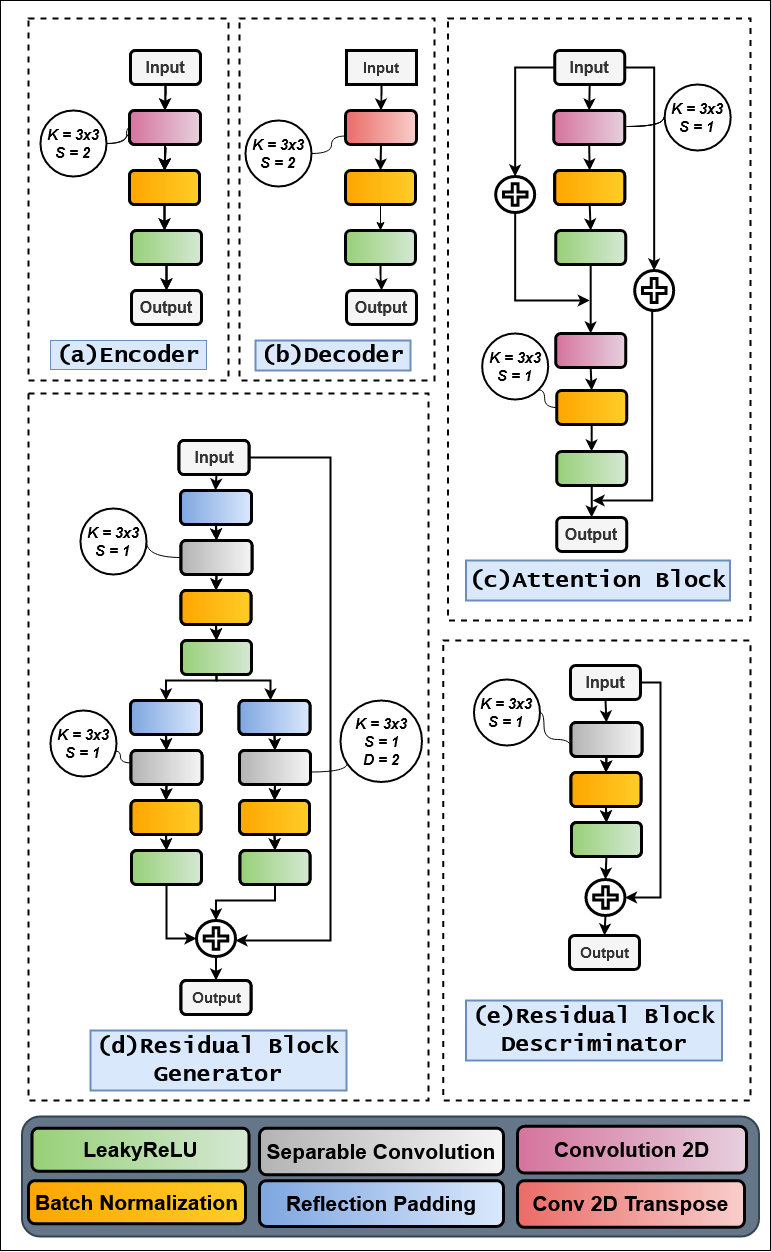}
    \caption{Individual blocks of our proposed GAN architecture consisting of (a) Encoder, (b) Decoder, (c) Attention block, (d) Residual Block for Generator and (e) Residual Block for Discriminator where K stands for kernel size, S is for stride  and D is for Dilation rate }\label{fig2}
\end{figure}

\subsection{Coarse and Fine Generators}
\label{subsec:generators}
Coupling coarse-to-fine generator for image translation tasks results in very pristine and high quality images , as witnessed in recent architectures, such as pix2pixHD \cite{wang2018high}, SPADE \cite{park2019semantic}, and Starganv2 \cite{choi2020stargan}. We incorporate this into technique in our architecture by using two generators ($G_{fine}$ and $G_{coarse}$), as illustrated in Fig.~\ref{fig1}. The generator $G_{fine}$ synthesizes smooth FA from fundus images by learning local information such as retinal venules, blood vessels, hemorrhages, exudates, and protein buildup. On the contrary, the generator $G_{coarse}$ tries to extract and preserve global information, such as the structures of the macula, optic disc, color intensity, contrast, and illumination, while producing less detailed angiograms. The generators consist of multiple encoders, decoder, attention, residual blocks, and a feature fusion block between the fine and coarse generator. $G_{fine}$  has an input dimension of  $512\times 512$ and produces outputs with the same resolution. Likewise, $G_{coarse}$ takes an image with half the resolution ($256\times 256$) and synthesizes an image with the same size. Additionally, the $G_{coarse}$ outputs a feature vector of the size $256\times 256 \times 64$ that is combined with one of the intermediate layers of $G_{fine}$ using the fusion operation. The representation of these generators is illustrated in Fig.~\ref{fig1}. In the following sections, we elaborate on each of these blocks in detail.

\subsection{Encoder and Decoder Blocks}
\label{subsec:encdec}
Both generators and discriminators incorporate the encoder blocks for downsampling the feature maps. On the contrary, only the generators use decoder blocks for upsampling to get the desired feature maps and output. The encoder block consists of a convolution layer followed by a batch-norm layer \cite{ioffe2015batch} and Leaky-ReLU activation function, as illustrated in Fig.~\ref{fig2}(a). In contrast, the decoder block comprises of transposed convolution layer and successive batch-norm \cite{ioffe2015batch} and Leaky-ReLU activation \ref{fig2}(b).  Interestingly, $G_{coarse}$ is downsampled twice ($\times 2$) using the encoder. After successive residual blocks, the decoder blocks are using to upsample twice again. For $G_{fine}$, the encoder is utilized once, and after the repetition of residual blocks, a single decoder is used to get the same spatial dimension of the output. We use a kernel size, $k=3$ and stride, $s=2$ for both of our convolution, and transposed convolution layers. 

\subsection{Distinct Residual Blocks for Generator and Discriminator}
\label{subsec:residualblock}
Lately, residual blocks have become the standard for generative models accomplishing image-to-image translation, image inpainting, and semantic segmentation tasks \cite{wang2018high,park2019semantic}. The fundamental design consists of a residual unit with two consecutive convolution layers and a skip connection that adds feature tensor of the input with the output. Regular convolution layers are computationally inefficient as opposed to separable convolution \cite{chollet2017xception}. Separable convolution consists of a depth-wise convolution followed by a point-wise convolution. By doing so, it extracts and retains the depth and spatial information through the network. Recent studies show that combining separable convolutional layers with dilation allows for more robust feature extraction~\cite{opticnet19}. We incorporate this technique to design two distinct novel residual blocks for our generators and discriminators, as shown in Fig.~\ref{fig2}(d) \& Fig.~\ref{fig2}(e). The residual block for our generator consists of a separable convolution layer followed by two branches of separable convolution. The difference is that one branch consists of a separable convolution with a dilation rate of, $d=1$ and the other with dilation rate, $d=2$. We use a kernel size, $k=3$ and stride, $s=1$ for all of our separable convolution layers. Each separable convolution is preceded by a Reflection padding layer and succeded by a Batch-Normalization and Leaky-ReLU activation layer. The skip connection and output of the two branches are all added together to produce the final output. In contrast, the residual block for the discriminator consists of a Separable convoluton layer, followed by Batch-Normalization and Leaky-ReLU activation function. The separable convolution has a kernel size of $k=3$ and stride, $s=1$.

\subsection{Attention block}
\label{subsec:attention}
Next, we elaborate our proposed attention block, as illustrated in
Fig.~\ref{fig2}(c). The block consists of two successive residual units, Convolution, BatchNorm, and Leaky-ReLU layers. Both convolution layer has kernel size of $k=3$ and stride, $s=1$. Other than that, there are two skip connections one coming from the input and being added to the output of the first residual unit. The other one is coming from the input and summed with the output of the last residual unit. We use attention block for combining feature information from the bottom layers of the network with the top layers of the network, as illustrated in Fig.~\ref{fig2}. $G_{coarse}$ comprises of two attention block, coming out of the with two encoders and being added with the two decoders successively. In contrast, the $G_{fine}$ has only one attention block between the encoder and decoder.  The reason behind utilizing attention block is to retrieve and retain spatial information that can be combined with the learned features of the later layers of the architectures as observed in similar GAN architectures \cite{zhang2019self,chen2018attention}.

\subsection{Multi-scale Markovian Discriminators}
\label{subsec:discriminators}
GAN discriminators need to adjust to coarse and fine generated outputs for distinguishing between real and synthesized images. To solve this underlying issue, we need a dense network with a huge amount of computable parameters. Alternatively, convolution with a wider receptive field can be utilized for extracting spatial information. This can easily lead to overfitting while training the model. To address this issue, we exploit the idea of using two Markovian discriminators, first introduced in a technique called PatchGAN \cite{li2016precomputed}. The method consists of discriminators with variable sized input resolution and can help with the overall adversarial training of the architecture as observed in \cite{wang2018high}

We use four discriminators that incorporates almost the same network structure but operate at two different resolutions. We organize the four discriminators into two sets, $D_{fine}=[D1_{f},D2_{f}]$ and $D_{coarse}=[D1_{c},D2_{c}]$ as illustrated in Fig.~\ref{fig1}.  We resize each of the coarse and fine angiograms and fundus with size $512\times512$ and $256\times256$ by a factor of $2$ using the Lanczos filter~\cite{duchon1979lanczos}. $D2_{f}$ and $D2_{c}$ have a unique average pooling layer right after the input which resizes the resolution to $256\times256$  and $128\times128$. Other than that, all four discriminators have identical layers consisting of three repetitive encoder and residual block pairs (in Fig.~\ref{fig2}(a) and Fig.~\ref{fig2}(e). Lastly, convolution layer is used for getting spatial dimension of $64\times64$ and $32\times32$  for $D1_{f},D2_{f}$ and $32\times32$ and $16\times16$ for $D1_{c},D2_{c}$  as outputs.

The coarse discriminators one that learns feature at a lower resolution tries to convince the coarse generator to retain more global features such as the macula, spherical optic disc, appearance, and illumination. On the other hand,  the fine discriminators dictate the fine generator to produce more detailed local features such as retinal vessels, arteries, exudates,  etc. By doing this we fuse features from both generators while training them autonomously with their joined multi-scale discriminators.

\subsection{Weighted Object Function and Adversarial Loss}
\label{subsec:objective}

With the given discriminators and generators, the objective function for our whole network can be formulated as Eq.~\ref{eq1}. It's a multi-objective problem of maximizing the loss of the discriminators while diminishing the loss of the generators. 
\begin{equation}
    \min \limits_{G_{f},G_{c}} \max \limits_{D_{f},D_{c}}  \mathcal{L}_{adv}(G_{f},G_{c}, D_{f},D_{c})
    \label{eq1}
\end{equation} 

For adversarial training, we use Hinge-Loss \cite{zhang2019self,lim2017geometric} as illustrated in Eq.~\ref{eq2} and Eq.~\ref{eq3}. Effectively, all the fundus images and their corresponding angiogram pairs, are normalized to $[-1,1]$. This in turn helps with widening the gap between the pixel intensities of the real and synthesized angio images. In Eq.~\ref{eq4} we add them and use $\lambda_{adv}$ as weight multiplier with the $\mathcal{L}_{adv}(G)$.
\begin{multline}
    \mathcal{L}_{adv}(D) = - \mathbb{E}_{x,y} \big[\ \min(0,-1+D(x,y))\big]\ \\-  \mathbb{E}_{x} \big[\ \min(0,-1-D(x,G(x))) \big]\
    \label{eq2}
\end{multline}
\begin{equation}
    \mathcal{L}_{adv}(G) = - \mathbb{E}_{x,y} \big[(D(G(x),y))\big]\
    \label{eq3}
\end{equation}
\begin{equation}
    \mathcal{L}_{adv}(G,D) = \mathcal{L}_{adv}(D) + \lambda_{adv} (\mathcal{L}_{adv}(G)) 
    \label{eq4}
\end{equation}
Here, In Eq.~\ref{eq2} and Eq.~\ref{eq3} the discriminators are first trained on the real fundus, $x$ and real angiogram, $y$, and then trained on the real fundus, $x$ and synthesized angiogram, $G(x)$. We begin by batch-wise training the discriminators $D1_{f}, D2_{f}$, and $D1_{c},D2_{c}$ for a couple of iterations on randomly sampled data. After that, we train the $G_{c}$ while keeping the weights of the discriminators frozen. In the same manner, we train the $G_{f}$ on a batch of random images while keeping weights of all the discriminators frozen. 

The generators also incorporate the reconstruction loss and perceptual loss~\cite{johnson2016perceptual} as shown in Eq.~\ref{eq5} and Eq.~\ref{eq6}. By utilizing these losses we ensure the synthesized images contain more realistic color, contrast, and vascular structure. We also employ feature matching loss ~\cite{wang2018high} with all our discriminators and as given in Eq.~\ref{eq7}.

\begin{equation}
    \mathcal{L}_{rec}(G) = \mathbb{E}_{x,y} \Vert G(x) - y \Vert^2
    \label{eq5}
\end{equation}

\begin{equation}
    \mathcal{L}_{perc}(G) = \mathbb{E}_{x,y} \sum_{i=1}^{k}\frac{1}{M} \Vert F_{vgg}^{i}(y) - F_{vgg}^{i}(G(x))\Vert
    \label{eq6}
\end{equation}

\begin{equation}
    \mathcal{L}_{fm}(G,D_{n}) = \mathbb{E}_{x,y} \sum_{i=1}^{k}\frac{1}{N} \Vert D_{n}^{i}(x,y) - D_{n}^{i}(x,G(x))\Vert
    \label{eq7}
\end{equation}

For Eq.~\ref{eq5}, $\mathcal{L}_{rec}$ is the reconstruction loss for a real angiogram, $y$, given a generated angiogram, $G(x)$. We use this loss for both $G_{f}$ and $G_{c}$ so that the model can generate high-quality angiograms of different scales. In Eq.~\ref{eq6}, $\mathcal{L}_{perc}$ calculates the difference between real and fake angio features extracted by pushing both of successively in VGG19 architecture~\cite{simonyan2014very}.  Lastly, Eq.~\ref{eq7} is calculated by taking the features from intermediate layers of the discriminator by first inserting the real and fake angiograms consecutively. Here, $M$ and $N$ stands for the number of feature layers extracted from VGG19 and the discriminators consecutively.

By incorporating Eq.~\ref{eq4}, \ref{eq5}, \ref{eq6} and \ref{eq7} we can formulate our final objective function as given in Eq.~\ref{eq8}.
\begin{multline}
\min \limits_{G_{f},G_{c}} \big( \max \limits_{D_{f},D_{c}}  (\mathcal{L}_{adv}(G_{f},G_{c}, D_{f},D_{c})) + \\ \lambda_{rec}\big[\ \mathcal{L}_{rec}(G_{f},G_{c})\big]\ + 
\lambda_{fm}\big[\ \mathcal{L}_{fm}(G_{f},G_{c}, D_{f},D_{c})\big]\ \\ + \lambda_{perc}\big[\ \mathcal{L}_{perc}(G_{f},G_{c})\big]\ \big)
\label{eq8}
\end{multline}

Here, $\lambda_{adv}$, $\lambda_{rec}$, $\lambda_{perc}$ and $\lambda_{fm}$ signifies loss weighting that are multiplied with their respective losses. The loss weighting dictates which networks to prioritize while training. For our architecture, more weight is given to the $\mathcal{L}_{adv}(G)$, $\mathcal{L}_{rec}$, $\mathcal{L}_{perc}$, and thus we select bigger $\lambda$ values for those. 

\section{Experiments}
In the next section, we detail our model experiments and evaluate our architecture based on qualitative and quantitative metrics. First, we elaborate on the structuring and pre-processing of our dataset in Sec.~\ref{subsec:dataset}. Then detail our hyper-parameter selection and tuning in Sec.~\ref{subsec:hyper}. Next, we describe our adversarial training scheme in Sec. ~\ref{subsec:training}. Also, we compare our architecture with existing state-of-the-art generative models based on some qualitative evaluation metrics in Sec.~\ref{subsec:quant}. Lastly, in Sec.~\ref{subsec:qual},  we analyze the quantification done by experts, by distinguishing between real and synthesized angiograms. 
\subsection{Dataset}
\label{subsec:dataset}
We use the fundus and angiography data-set provided in \cite{hajeb2012diabetic}. It consists of thirty image and twenty-nine pairs of the healthy and unhealthy fundus and angiogram images, collected from fifty-nine individual patients. Next, we clean the dataset by taking only seventeen pairs of images based on one-to-one alignment between the fundus and angiogram pairs. These image pairs are either accurately aligned or almost aligned. The original image size is $576\times720$, but we take 50 overlapping crops of $512\times512$  sized samples from each. By doing so, we end up having  850 images in total for training. The fundus images are in RGB format, and angiograms are in a Gray-scale format. For testing, we take fourteen image pairs and crop four overlapping quadrants of the image to generate a test set of fifty-six test images.

\begin{algorithm}[h]
\caption{Attention2AngioGAN training}
\label{alg1}
\begin{algorithmic}[1]
 \renewcommand{\algorithmicrequire}{\textbf{Input:}}
 \renewcommand{\algorithmicensure}{\textbf{Output:}}
  \REQUIRE $x_{i} \epsilon X$, $y_{i} \epsilon {Y}$
 \ENSURE $G_{f}$, $G_{C}$
  \STATE \textbf{Initialize hyper-parameters}: \\$max\_epoch$, $b$, $max\ d\_iter$, $\omega^f_{D}$, $\omega^c_{D}$, $\omega^f_{G}$, $\omega^c_{G}$  $\alpha^f_{D}$, $\alpha^c_{D}$, \\  $\alpha^f_{G}$, $\alpha^c_{G}$, $\beta^f_{D}$, $\beta^c_{D}$, $\beta^f_{G}$,  $\beta^c_{G}$, $\lambda_{rec}$, $\lambda_{per}$, $\lambda_{fm}$, $\lambda_{adv}$ 
  \FOR{$e=0\ to\ max\_epoch$}
  \STATE Sample $x_{f},x_{c},y_{f},y_{c}$, using batch-size $b$ 
    \FOR{$d\_iter=0\ to\ max\ d\_iter$}
        \STATE $\mathcal{L}_{adv}(D) \gets D_{c}(x_{c},y_{c})$, $D_{f}(x_{f},y_{f})$  
        \STATE $\mathcal{L}_{adv}(D) \gets D_{c}(x_{c},G_{c}(x_{c}))$, $D_{f}(x_{f},G_{F}(x_{f}))$  
        \STATE $\mathcal{L}_{adv}(G) \gets G_{c}(x_{c})$, $G_{F}(x_{f})$
        \STATE $\mathcal{L}_{adv}(G,D) \gets \mathcal{L}_{adv}(D) + \lambda_{adv}(\mathcal{L}_{adv}(G))$
        \STATE $\omega^c_{D} \gets \omega^c_{D}+ Adam(D_c,G_c,\omega^c_{D},\alpha^c_{D},\beta^c_{D})$
         \STATE $\omega^f_{D} \gets \omega^f_{D}+ Adam(D_f,G_f,\omega^f_{D},\alpha^f_{D},\beta^f_{D})$
    \ENDFOR
    \item[] \textbf{Freeze $\omega^c_{D},\omega^f_{D}$}
    \STATE Sample $x_{f},x_{c},y_{f},y_{c}$, using batch-size $b$
    \STATE $\mathcal{L}_{rec}(G_c) \gets G(x_{c}), y_{c}$,
    \STATE $\mathcal{L}_{rec}(G_f) \gets G(x_{f}),y_{f}$
    \STATE $\mathcal{L}_{perc}(G_c) \gets F_{vgg}^{c}(y), F_{vgg}^{c}(G(x))$
    \STATE $\mathcal{L}_{perc}(G_f) \gets F_{vgg}^{f}(y), F_{vgg}^{f}(G(x))$
    \STATE $\omega^c_{G} \gets \omega^c_{G}+ Adam(G_c,\omega^c_{G},\alpha^c_{G},\beta^c_{G})$
    \STATE $\omega^f_{G} \gets \omega^f_{G}+ Adam(G_f,\omega^f_{G},\alpha^f_{G},\beta^f_{G})$
    \item[] \textbf{Unfreeze $\omega^c_{D},\omega^f_{D}$}
    \STATE $\mathcal{L}_{fm}(D_c) \gets D^c_n(x_{c},y_{c}) , D^f_n(x_{c},G(x_{c})) $  
    \STATE $\mathcal{L}_{fm}(D_f) \gets D^f_n(x_{f},y_{f}) , D^f_n(x_{f},G(x_{f})) $
    \STATE $\omega^c_{D} \gets \omega^c_{D}+ Adam(D_c,G_c,\omega^c_{D},\alpha^c_{D},\beta^c_{D})$
    \STATE $\omega^f_{D} \gets \omega^f_{D}+ Adam(D_f,G_f,\omega^f_{D},\alpha^f_{D},\beta^f_{D})$
    \item[] \textbf{Freeze $\omega^c_{D},\omega^f_{D}$}
    \STATE $\mathcal{L}_{adv}(D) \gets D_{c}(x_{c},y_{c})$, $D_{f}(x_{f},y_{f})$  
    \STATE $\mathcal{L}_{adv}(D) \gets D_{c}(x_{c},G_{c}(x_{c}))$, $D_{f}(x_{f},G_{F}(x_{f}))$  
    \STATE $\mathcal{L}_{adv}(G) \gets G_{c}(x_{c})$, $G_{F}(x_{f})$
    \STATE $\mathcal{L}_{adv}(G,D) \gets \mathcal{L}_{adv}(D) + \lambda_{adv}(\mathcal{L}_{adv}(G))$
    \STATE $\omega^c_{D} \gets \omega^c_{D}+ Adam(D_c,G_c,\omega^c_{D},\alpha^c_{D},\beta^c_{D})$
    \STATE $\omega^f_{D} \gets \omega^f_{D}+ Adam(D_f,G_f,\omega^f_{D},\alpha^f_{D},\beta^f_{D})$
    \STATE Save weights and snapshot of $G_{f},G_{c}$
    \STATE $\mathcal{L} \gets \mathcal{L}_{adv} +  \lambda_{rec}(\mathcal{L}_{rec}) + \lambda_{fm}(\mathcal{L}_{fm}) + \lambda_{perc}(\mathcal{L}_{perc}) $
  \ENDFOR
\end{algorithmic}
\end{algorithm}

\subsection{Hyper-parameter tuning}
\label{subsec:hyper}
For adversarial training, we used hinge loss \cite{zhang2019self,lim2017geometric}. We picked $\lambda_{adv}=10$ (Eq.~\ref{eq4}) and $ \lambda_{rec} =10$, $ \lambda_{perc} =10$, $ \lambda_{fm} =1$ (Eq.~\ref{eq8}). For optimizer, we used Adam \cite{kingma2014adam}, with learning rate $\alpha=0.0002$, $\beta_1=0.5$ and $\beta_2=0.999$. We train with mini-batches with batch size, $b=2$ for 200 epochs. It took approximately 48 hours to train our model on NVIDIA P100 GPU.

\subsection{Training procedure}
\label{subsec:training}
In this section, we elaborate on our detailed algorithm provided in Algorithm~\ref{alg1}. To train our Attention2AngioGAN, we start by initializing all the hyper-parameters. Next, we sample a batch of the real fundus and angiogram $x,y$ images. We train the real fundus and fake angiogram pairs with $D_{f}, D_{c}$. After that, we use $G_{f}, G_{c}$ to synthesize fake angiograms and use the real fundus and fake angiograms, $x,G(x)$ to train discriminators $D_{f},D_{c}$. Following that, we calculate the adversarial loss, $\mathcal{L}_{adv}(D, G)$, and update the weights. We freeze the weights of the discriminators. Next, we train the generators and calculate the $\mathcal{L}_{rec}(G)$, $\mathcal{L}_{perc}(G)$ losses and update both generator's weights.  Subsequently, we unfreeze both discriminators weights, calculate the feature matching loss $\mathcal{L}_{fm}(D)$  and update the discriminator weights. In the final stage, we freeze both discriminator's weights and jointly fine-tune all the discriminator and generators. We calculate the total loss by adding and multiplying with their relative weights. For testing, we save the snapshot of the model and its weights.

\subsection{Qualitative Evaluation}
\label{subsec:qual}
For assessing the performance of our network, we used 14 test samples and cropped four quadrants of the image with a size of $512\times512$. We conducted two sets of tests to evaluate our networks. First, for estimating the accurate visual representation without transforming the image. Next, for global and local changes due to transformation and distortion of the image. By doing so, we measured the network's ability to adjust to structural changes to the vascular patterns and formation of the retinal subspace. We used the GNU Image Manipulation Program (GIMP) \cite{gimp2019gimp} for carrying out transformation and distortion on the images.
 
\begin{figure}[ht]
    \centering
    \includegraphics[width=8cm,height=14cm]{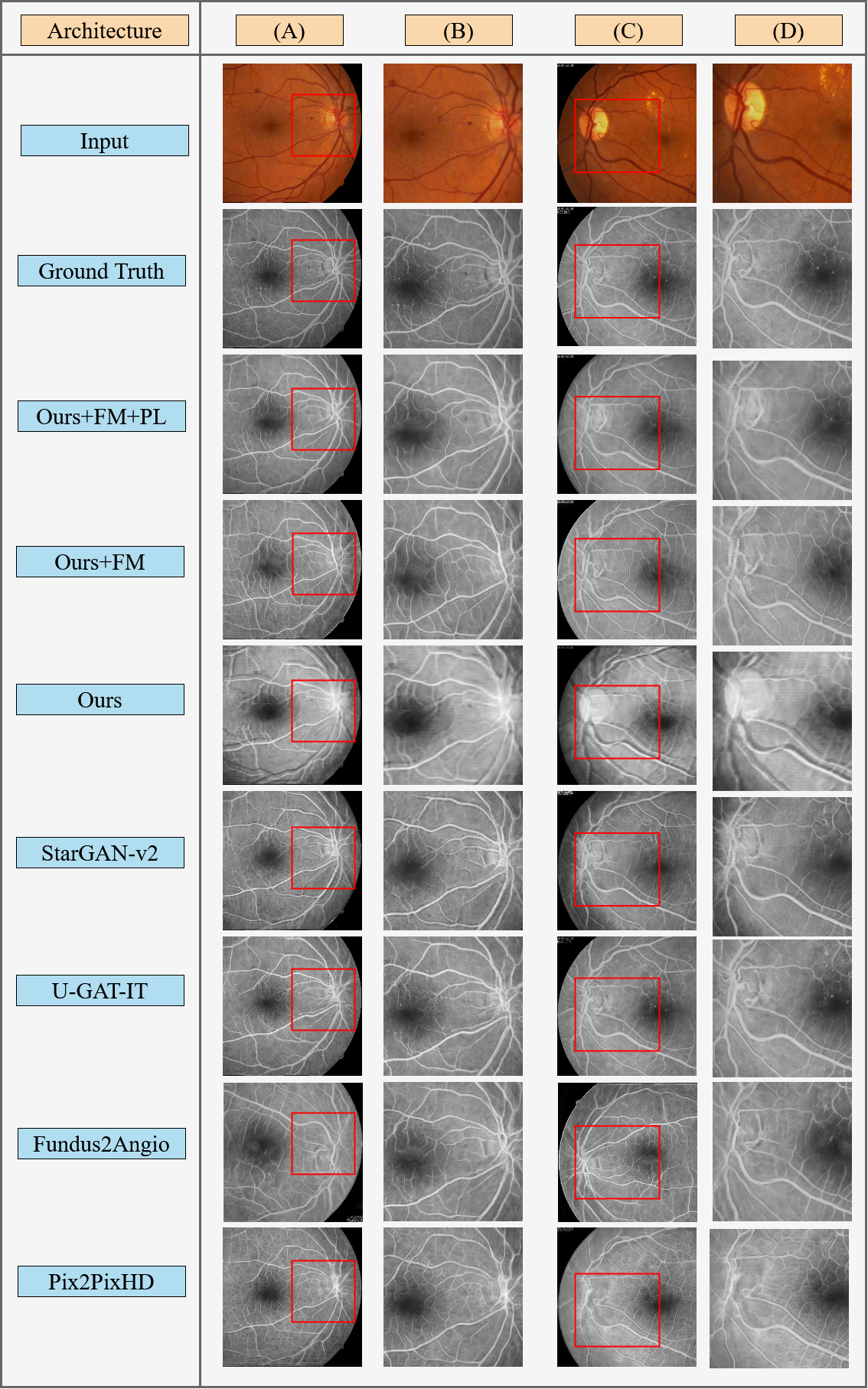}
    \caption{Comparativie results of different Angiograms generated using different state-of-the-art architectures. Column (A) and (C) represents two samples of real fundus, real angio and predicted angio images. Whereas column (B) and (D) represents the red rectangle box to show zoomed in local venular structures corresponding.}
    \label{fig3}
\end{figure}

\begin{figure}[ht]
    \centering
    \includegraphics[width=9cm,height=14cm]{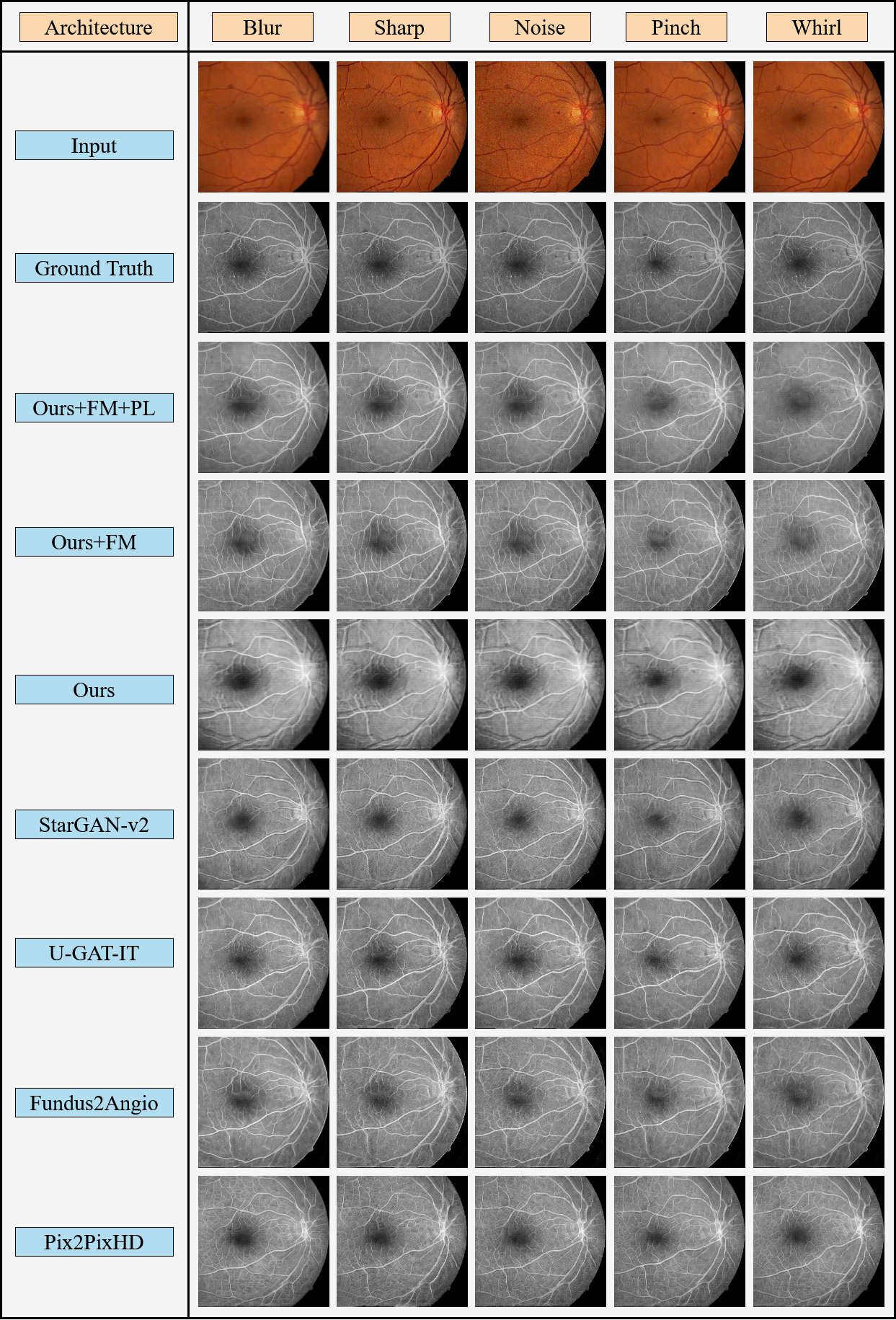}
    \caption{Angiogram generated from transformed and distorted Fundus images with natural changes, imaging errors and biological markers.}
    \label{fig4}
\end{figure}

For the first experiment, we train three variants of our network and four other state-of-the-art image-to-image translation architecture using the same number of epochs and batches of images. Next, we evaluate them using the same test sample. A side by side comparison of the results is illustrated in Fig.~\ref{fig3}. Column A \& C in Fig.~\ref{fig3} shows the global changes while column B \& D are zoomed-in to display local vascular structure and other local information. We use FM and PL to denote feature-matching and perceptual loss. By the looks of it, both of our models using the FM loss (with or without PL) produces vivid and convincing results. On the other hand, the visual result of our network without FM and PL produces distorted local structure due to not learning contrast and color of the optic disc using perceptual loss. Fundus2Angio and StarGANv2 also produce impressive results. However, if seen in the closed-up versions in columns B \& D, we can witness the right upper portion of the optic discs contains fewer blood vessels compared to our ones. U-GAT-IT and Pix2PixHD also fail to generate rich venules,  exudates, and protein buildup. 

In the second set of experiments, three transformations and two distortions were applied on the fundus images: 1) blurring to represent severe cataracts, 2) sharpening to represent dilated pupils, 3) signal noise to represent machine impedance during fundoscopy, 4) pinch, to visualize the pulled/pushed vascular formation, 5) whirl, for representing distortions caused by increased intraocular pressure (IOP). Improved robustness and adaption are represented by the generated angiograms similarity to the real FA image since these transformations and distortions may or may not affect the vascular structure of the retina. A side by side comparison of different architecture's predictions on these transformed images is illustrated in Fig.~\ref{fig4}. As it can be observed from the image, the proposed architecture produces images very similar to the ground-truth (GT). The results under these global and local vascular changes applied to the fundus image.

\begin{table*}[ht]
\centering

\caption{Test results for different architectures}
\begin{adjustbox}{width=5.8in,totalheight=2.8in}
\begin{threeparttable}
    \begin{tabular}{|l|c|c|c|c|c|c|} 
    \hline
    \multicolumn{7}{|c|}{\textbf{Fréchet Inception Distance (FID)}}\\
    \hline
    Architecture &  Orig. &  Noise & Blur &  Sharp & Whirl & Pinch \\
    \hline
    \textbf{Ours + PL\tnote{1} + FM\tnote{2}} & 24.6 &  21.6 (3.0$\downarrow$) &  30.0 (5.4$\uparrow$) &  25.6 (1.0$\uparrow$) &  40.0 (15.4$\uparrow$) &  24.9 (0.3$\uparrow$) \\
    \textbf{Ours + FM\tnote{2}} & \textbf{20.7} &  \textbf{20.8 }(0.1$\uparrow$) &  \textbf{23.5} (2.8$\uparrow$) &  \textbf{24.9} (4.2$\uparrow$) & \textbf{27.8} (7.1$\uparrow$) &  \textbf{19.5} (1.2$\downarrow$) \\
    \textbf{Ours} & 47.5 &  43.1 (4.4$\downarrow$) &  49.8 (2.3$\uparrow$) &  50.5 (3.5$\uparrow$) &  61.9 (14.5$\uparrow$) &  46.7 (0.8$\downarrow$) \\ 
    StarGAN-v2 \cite{choi2020stargan} & 27.7 & 35.1 (7.4$\uparrow$) & 32.6 (4.9$\uparrow$) & 27.4 (0.3$\downarrow$) & 32.7 (5.0$\uparrow$) & 26.7 (1.0$\downarrow$) \\
    U-GAT-IT \cite{kim2019u} & 24.5 & 26.0 (1.5$\uparrow$) & 30.4 (5.9$\uparrow$)  & 26.8 (2.3$\uparrow$) & 33.0 (9.5$\uparrow$) & 29.1 (4.6$\uparrow$) \\
    Fundus2Angio \cite{kamran2020fundus2angio} & 30.3 &  41.5 (11.2$\uparrow$) & 32.3 (2.0$\uparrow$) & 34.3 (4.0$\uparrow$) & 38.2 (7.9$\uparrow$) & 33.1 (2.8$\uparrow$) \\ 
    Pix2PixHD \cite{wang2018high} & 42.8  & 53.0 (10.2$\uparrow$)& 43.7 (1.1$\uparrow$) & 47.5 (4.7$\uparrow$) & 45.9 (3.1$\uparrow$) & 39.2 (3.6$\downarrow$) \\ 
    \hline
    \hline
    \multicolumn{7}{|c|}{\textbf{Kernel Inception Distance (KID)}}\\
    \hline
    Architecture &  Orig. &  Noise & Blur &  Sharp & Whirl & Pinch \\
    \hline
    \textbf{Ours + PL\tnote{1} + FM\tnote{2}} & \textbf{0.00087} & \textbf{0.05045} & \textbf{0.00235} & \textbf{0.05162} & 0.05390 & \textbf{0.04575} \\
    \textbf{Ours + FM\tnote{2}} & 0.00392 &	0.05390 & 0.00505 & 0.05301 & 0.05657	& 0.05341 \\
    \textbf{Ours} & 0.00595 & 0.05237 &	0.00617 & 0.05298 &	0.05613 & 0.05419 \\
    StarGAN-v2 \cite{choi2020stargan} & 0.00118 & 0.05274 & \textbf{0.00235} & 0.05331 & 0.05539 & 0.05271 \\
    U-GAT-IT \cite{kim2019u} & 0.00131	& 0.05610 & 0.00278 & 0.05533 & 0.05815 & 0.05719\\
    Fundus2Angio \cite{kamran2020fundus2angio} & 0.00184 & 0.05328 & 0.00272 & 0.05267 & \textbf{0.05278} & 0.04985 \\ 
    Pix2PixHD \cite{wang2018high} & 0.00258 & 0.05613 & 0.00254 & 0.05788 & 0.06029 & 0.05838 \\ 
    \hline
    \end{tabular}
    \begin{tablenotes}
         \item[1] PL = Perceptual Loss; FM = Feature-Matching Loss
         \item[2] FID: Lower is better; KID: Lower is better
    \end{tablenotes}
\end{threeparttable}
\end{adjustbox}
\label{table1}
\end{table*}

In the case of \textbf{blurred} fundus images, our architecture with and without PL, is less affected compared to other state-of-the-art models, as seen in (row 6 to 9 of column 1) of Fig.~\ref{fig4}. The venular and cellular structures are better conserved as opposed to StarGANv2 and Pix2PixHD. For \textbf{sharpened} fundus, the angiogram produced by UGATIT and Fundus2Angio (row 7 and 8 of column 2) exhibits grainy artifacts around the blood vessels, which are not present in our model with and without PL. For \textbf{noisy} images, our result with and without PL, is still unaffected with this pixel-level modification. However, all other state-of-the-art models (row 6 to 9 of column 3) fail to generate thin and small venular formations by failing to extract local features from the retinal subspace. On the contrary, our model without PL and FM produces jittery motion artifacts and high contrast around the border of the optic disc for all these transformations.

For distortions like \textbf{Pinch} and \textbf{Whirl}, our experimental result with and without perceptual loss shows the versatility and reproducibility of the proposed network to uncover the changes in vascular structure as seen in Fig.~\ref{fig4} (row 3 and 4 of column 4 and 5). Compared to ours, only StarGANv2 and U-GAT-IT maintains the flattening condition and manifestation of vascular changes but loses the overall smoothness in the process (row 6 to 7 of column 4 and 5). As seen in Fig.~\ref{fig4} ours with and without PL network encodes the feature information of vessel structures and is much less affected by both kinds of contortion. The other architectures failed to generate microvessel structure due to IOP or vitreous changes as can be seen in Fig.~\ref{fig4}.Contrarily, our model without any perceptual and feature-matching loss fails to encode this information and vascular changes. Consequently, For all kinds of transformation and distortion our model with and without perceptual triumphs over existing state-of-the-art image-to-image translation models.

\begin{table}[ht!b]
\caption{Results of Qualitative with Undisclosed Portion of Fake/Real Experiment}
\label{table2}
\centering
\begin{adjustbox}{width=3.4in,totalheight=0.9in}
\begin{threeparttable}
\begin{tabular}{|l|l|c|c|c|c|c|} 
\hline
&&\multicolumn{2}{c|}{\small Results} & \multicolumn{3}{c|}{ \small Average}  \\\hline
&& \small Correct & \small Incorrect & \small Missed\tnote{1} & \small Found\tnote{1} & \small Precision\tnote{2}\\\hline\hline
\small \multirow{2}{*}{Ours + FM + PL} &\small Fake & \small 10\% & \small 90\% & \small \multirow{2}{*}{55\%} & \small \multirow{2}{*}{45\%} & \multirow{2}{*}{\small \textbf{47.1\%}} \\
&\small Real & \small 80\% & \small 20\% & & & \\
\hline
\small \multirow{2}{*}{Ours + FM} &\small Fake & \small 12\% & \small 88\% & \small \multirow{2}{*}{53\%} & \small \multirow{2}{*}{47\%} & \multirow{2}{*}{\small \textbf{48.2\%}} \\
&\small Real & \small 82\% & \small 18\% & & & \\
\hline
\end{tabular}
    \begin{tablenotes}
         \item[1] Missed higher is better; Found lower is better
         \item[2] Precision Lower is better
    \end{tablenotes}
\end{threeparttable}
\end{adjustbox}
\end{table}
\subsection{Quantitative Evaluations}
\label{subsec:quant}
For quantitative evaluation, we performed two experiments. In the first experiment we use the Fréchet inception distance (FID) \cite{heusel2017gans} and Kernel Inception distance (KID) \cite{binkowski2018demystifying} which has been used to evaluate similar style-transfer GANs \cite{choi2020stargan,kim2019u}. We computed the FID and KID scores for different architectures on the generated FA image and original angiogram, including the five transformations and distortions. The results are reported in Table.~\ref{table1}. It should be noted that lower FID and KID score means better results. 

From Table.~\ref{table1},  out of our three networks, the best FID is achieved for ours without PL.  And it achieves the lowest scores among out of all other architecture, for both with and without distortions. For KID, our model with PL achieves the lowest score for four out of five types of distortions. Fundus2Angio scores lower KID for distorted images using whirl. Other than that, StarGAN-v2 achieves the same score as our network having a KID of 0.00235.

In the next experiment, we assess the quality of the synthesized angios by asking three expert ophthalmologists to identify fake angios among a collection of 50 balanced (50\%, 50\%) and randomly set of mixed angiograms. For this experiment, the exact number of fake and real images was not known by the experts. By not disclosing this information we tried to evaluate following criterion: 1) Correct fake and real angios found by the experts, where lower is better, 2) Incorrect fake and real angios missed by the experts, where higher is better and 3) The average precision representing the effective the identification is by the experts, where lower is better. The detailed results are shown in Table~\ref{table2}.

As it can be seen from Table~\ref{table2}, experts assigned 90\% and of 88\% of the fake angiograms as real, for images generated by two of our models. The result also shows that experts had difficulty in identifying fake images, while they easily identified real angiograms with 80\% and 82\% certainty. On average, the experts misclassified 55\% and 53\% of all images for two of our models consecutively. The average precision diagnosis of the experts are 47.1\% and 48.2\%. Consequently, our model with lower precision achieves the best result by fooling the experts to identify fake angios as real.

\section{Conclusion}
In this paper, we proposed a new image-to-image translation architecture called Attention2AngioGAN. The architecture synthesizes high quality and vivid looking angiograms from fundus images without any expert intervention. Additionally, we illustrated its robustness, flexibility, and reproducibility by producing high-quality angiograms from transformed and distorted images, which imitates biological markers seen in real fundus images. As a result, the proposed network can be efficiently employed to generate precise FA images of patients developing disease overtime. This is best suited for disease progression monitoring to predict the development of diseases in vivo. We hope to extend this work for other areas of ophthalmological data modalities.

\bibliographystyle{IEEEtran}
\bibliography{reference}


\end{document}